\documentclass[fractalfract,article,submit,moreauthors,pdftex,10pt,a4paper]{mdpi} 

\firstpage{1} 
\articlenumber{x}
\doinum{10.3390/------}
\pubvolume{xx}
\pubyear{2016}
\copyrightyear{2016}
\externaleditor{Academic Editor: name}
\history{Received: date; Accepted: date; Published: date}
 \theoremstyle{mdpi}
 \newcounter{thm}
 \newcounter{ex}
 \newcounter{re}
 \newcommand{\res}{\mathrm{Res}}
\newcommand{\id}{\mathrm{d}}
\newcommand{\ud}{\mathrm{d}}

\Title{Option Pricing Models Driven by the Space-Time Fractional Diffusion: Series Representation and Applications}
\Author{Jean-Philippe Aguilar$^{1,\star}$, Jan Korbel$^{2,3,4}$}
\AuthorNames{Jean-Philippe Aguilar, Jan Korbel}

\address{%
$^{1}$ \quad BRED Banque Populaire, Modeling Department, 18 quai de la Râpée, Paris - 75012; jean-philippe.aguilar@bred.fr\\
$^{2}$ \quad Section for the Science of Complex Systems, CeMSIIS, Medical
University of Vienna, Spitalgasse 23, A-1090, Vienna, Austria\\
$^{3}$ \quad  Complexity Science Hub Vienna, Josefst\"{a}dterstrasse 39, 1080 Vienna, Austria\\
$^{4}$ \quad Faculty of Nuclear Sciences and Physical Engineering, Czech Technical University in Prague; korbeja2@fjfi.cvut.cz}

\corres{Correspondence: jean-philippe.aguilar@bred.fr}%; Tel.: +x-xxx-xxx-xxxx}

\abstract{In this paper, we focus on option pricing models based on space-time fractional diffusion. We briefly revise recent results which show that the option price can be represented in the terms of rapidly converging double-series and apply these results to the data from real markets. We focus on estimation of model parameters from the market data and estimation of implied volatility within the space-time fractional option pricing models.}

\keyword{Space-time fractional diffusion, European option pricing, Mellin transform, Multidimensional complex analysis}

\begin{document}

%%%%%%%%%%%%%%%%%%%%%%%%%%%%%%%%%%%%%%%%%%
%% Sections that are not mandatory are listed as such. The section titles given are for Articles. Review papers and other article types have a more flexible structure. 

%% Only for the journal Gels: Please place the Experimental Section after the Conclusions

%%%%%%%%%%%%%%%%%%%%%%%%%%%%%%%%%%%%%%%%%%
\section{Introduction}

The pricing of derivatives, and notably of options, is a central subject in mathematical finance. It allows the market practitioner to estimate the value of its portfolio, and to construct appropriate hedging strategies. The most popular option pricing model is the one introduced by Black and Scholes \cite{BS73}, because of its simplicity (e.g., the option price can be expressed in terms of simple mathematical functions), and can be used to imply the market parameters, such as volatility surfaces, from the observation of traded market prices.

On the other hand, the simplicity of the Black-Scholes (BS) model is also its main limitation. The dynamics of the underlying asset is described by a geometric Brownian motion, so the resulting option price is given by the Gaussian distribution. This assumption does not describe well extreme market events such as sudden jumps of asset prices, which happen far more frequently than expected in the Gaussian world (see for instance the influential book by Taleb \cite{Taleb10}). This makes the Black-Scholes formula less reliable in abnormal conditions or in illiquid markets.

During the past years, several models have been introduced to describe the market dynamics more realistically; one can mention regime switching multifractal models \cite{Calvet08}, stochastic volatility models \cite{Heston93} or jump (L\'evy-stable) processes \cite{Tankov03}. More recently, a model based on space-time fractional diffusion has been introduced \cite{KK16,Korbel16}, and can be regarded as a generalization of the L\'evy-stable model; analytic resolution of this model has been provided in \cite{ACK17} under the form of a series representation for its pricing formula. In the present article, we briefly recall these analytic results (and notably how they recover previously known models) and test their efficiency in real market applications. We also discuss the related topics as at-the-money approximation or implied volatility.

The paper is organized as follows. In the following section, we introduce some fundamental concepts in option pricing, and the main models that, under risk-neutral approach, can be reduced to a space-time fractional diffusion problem. This includes the BS model (which reduces to the classical heat equation), the L\'evy-stable model (which reduces to the space-fractional diffusion equation) and the generic space-time (or double) fractional model. In section 3, we briefly recall the analytic solution to the space-time fractional model. In section 4 we present various applications of the theoretical results: we calculate the call prices and compare it with the real data, we introduce at-the-money volatility and discuss the construction of volatility smile.  The last section is dedicated to conclusions.

\section{Option pricing}

The price of an option of strike $K$ and maturity $T$, is a function of market parameters such as an underlying asset price $S$, a risk-free interest rate $r$ and a market volatility $\sigma$. We will denote this price by $V(S,K,r,\sigma,t)$. In the case of an European option, it is characterized by its payoff, that its, its value at the exercise time $T$; for an European call, this value is equal to 
\begin{equation}
V(S,K,r,\sigma,T) \, = \, \max \{ S-K, 0 \} \, := \, [S-K]^+
\end{equation}
For a  put option, the corresponding payoff is $[K-S]^+$. Now we recall the principles of option pricing, that is, the way of determining $V(S,K,r,\sigma,t)$.

\subsection{The risk-neutral approach}

The risk-neutral, or risk-free approach is based on the idea that one can construct a portfolio where the (market) risk can be totally eliminated \cite{Wilmott06}. Schematically, it consists in buying an option and selling a certain quantity $\Delta$ (to be determined) of the underlying price, so that the total value of the portfolio reads $\Pi \, = \, V - \Delta S$ and therefore:
\begin{equation}\label{RiskNeutral1}
\ud \Pi \, = \, \ud V - \Delta \ud S
\end{equation}
On the other hand, the markets are assumed to offer no arbitrage opportunity, that is, any risk-less portfolio will have the same yield as if it were capitalized at the risk-free interest rate:
\begin{equation}\label{RiskNeutral2}
\ud \Pi \, = \, r\Pi \ud t
\end{equation}
Equalizing \eqref{RiskNeutral1} and \eqref{RiskNeutral2} and making an appropriate choice for $\Delta$ transforms the option pricing problem into the resolution of a partial differential equation with terminal condition.

\noindent
From a more theoretical point of view, within risk-neutral approach, the price can be formulated as the discounted expectations of the terminal payoff \cite{Privault14}:
\begin{equation}
V(S,K,r,\mu,\tau) \, = \, e^{-r\tau} \, \mathbb{E}^{\mathbb{Q}} \left[ [S-K]^+ \right]
\end{equation}
where we have introduced the time-to-maturity $\tau = T-t$. The expectations are to be taken under the risk-neutral measure $\mathbb{Q}$, which is associated to the original probability measure $\mathbb{P}$ via the Radon-Nikodym derivative:
\begin{equation}
\frac{\ud \mathbb{Q}_t}{\ud \mathbb{P}_t} \, = \, e^{S_t - \mu t}
\end{equation}
The risk-neutral parameter $\mu$ can be expressed as
\begin{equation}\label{mu_def}
\mu \, = \, - \log \mathbb{E}^{\mathbb{P}} \left[ e^{S_{t=1}} \right]
\end{equation}
More details can be found in \cite{Gerber93,KK16}.

\subsection{Black-Scholes model}
In the BS model, the underlying asset price $S$ is assumed to be described by a geometric Brownian motion:
\begin{equation}\label{BS_SDE}
\ud S_t \,= \, r \, S \ud t\, + \, \sigma \, S \, \ud W_t
\end{equation}
It follows from Itô's lemma \cite{Oksendal00} that the total differential of the option price is:
\begin{equation}
\ud V \, = \, \frac{\partial V}{\partial t} \,\ud t \, + \, \frac{\partial V}{\partial S} \,\ud S \, + \, \frac{1}{2} \sigma^2 S^2 \frac{\partial^2 V}{\partial S^2} \, \ud t
\end{equation}
Choosing $\Delta=\frac{\partial V}{\partial S}$, using \eqref{BS_SDE} and equalizing \eqref{RiskNeutral1} and \eqref{RiskNeutral2}, we have shown that the call price satisfies the famous Black-Scholes equation, which is a partial differential equation (PDE) with terminal condition:
\begin{align}\label{BS_Equation}
\left\{
\begin{aligned}
 & \frac{\partial V}{\partial t} \, + \, \frac{1}{2} \sigma^2 S^2\frac{\partial^2 V}{\partial S^2} \, + \, r S \frac{\partial V}{\partial S} \, - \, rV \, = \, 0  \hspace*{1cm} t\in[0,T]  \\
 & V(S,K,r,\sigma,t=T) \, = \, [S-K]^+
\end{aligned}
\right.
\end{align}
It is known that, with the change of variables
\begin{align}
\left\{
\begin{aligned}
 & x \, := \, \log S \, + \, (r-\frac{\sigma^2}{2}) \, \tau \\
 & \tau \, := \, T-t \\
 & V(S,K,r,\sigma,t)  \, := \, e^{-r\tau}W(x,K,r,\sigma,\tau)
\end{aligned}
\right.
\end{align}
then the Black-Scholes PDE \eqref{BS_Equation} resumes to the diffusion (or heat) equation
\begin{equation}\label{heat}
\frac{\partial W}{\partial \tau} \, - \, \frac{\sigma^2}{2}\frac{\partial^2 W}{\partial x^2} \, = \, 0
\end{equation}
which is a particular case of the double fractional diffusion \eqref{double_fractional} with time fractionality $\gamma=1$ and space fractionality $\alpha = 2$. It is well known 
that the Green function for \eqref{heat} is the heat kernel
\begin{equation}
g(x,K,r,\sigma,\tau) \, = \, \frac{1}{\sigma\sqrt{2\pi\tau}} \, e^{-\frac{x^2}{2\sigma^2\tau}}
\end{equation}
and therefore, by the method of Green functions and turning back to the initial variables, we obtain the solution for the Black-Scholes PDE (in the call case):
\begin{equation}\label{BS_GreenForm}
V(S,K,r,\sigma,\tau) \, = \, e^{-r\tau} \, \int\limits_{-\infty}^{+\infty} [Se^{(r-\frac{\sigma^2}{2})\tau+y}-K]^{+} \, g(y,K,r,\sigma,\tau) \, \id y
\end{equation}
Basic manipulations on the integral \eqref{BS_GreenForm} yield
\begin{equation}\label{BS_Formula}
V(S,K,r,\sigma,\tau) \, = \, SN(d_+) - Ke^{-r\tau} N(d_-) \hspace*{1cm} d_{\pm} = \frac{1}{\sigma\sqrt{\tau}}\left(\log\frac{S}{K}+r\tau \right) \pm \frac{1}{2}\sigma\sqrt{\tau}
\end{equation}
where $N(.)$ is the normal distribution function; formula \eqref{BS_Formula} is the celebrated \textit{Black-Scholes formula} for the European call. The corresponding risk-neutral parameter is therefore 
\begin{equation}
\mu_{BS} = - \frac{\sigma^2}{2}
\end{equation}

\subsection{Finite-Moment L\'evy-stable model}

An interesting generalization of the BS model is the so-called Finite Moment L\'evy (or Log) Stable (FMLS) model; it was introduced in \cite{Carr03} and assumes that the underlying asset price $S_t$ is described by:
\begin{equation}
\ud S_t \, = \, r S_t \ud t \, + \, \sigma S_t \ud L_{\alpha,\beta}(t)
\end{equation}
where $L_{\alpha,\beta}(t)$ is the L\'evy process \cite{Zolotarev86}. $\alpha\in[0,2]$ and $\beta\in[-1,1]$ are the so-called \textit{stability} and \textit{asymmetry} parameters and determine the decay of the tails and the asymmetry of the probability distributions $g_{\alpha,\beta}(x,t)$. Under the (strong) hypothesis that $\beta=-1$ (maximal negative asymmetry hypothesis) then the distribution $g_{\alpha,-1}:=g_{\alpha}$ possesses one heavy-tail in the negative axis, and another tail in the positive axis with exponential decay as soon as $\alpha>1$, and finite exponential moments 
\begin{equation}\label{Levy_moment}
\mathbb{E}^\mathbb{P} \, \left[ e^{- \lambda S_{t}} \right] \, = \, e^{-\lambda^\alpha  \frac{\left( \frac{\sigma}{\sqrt{2}} \right)^\alpha}{\cos\frac{\pi\alpha}{2}} }
\end{equation}

\noindent
These particular L\'evy distributions are sometimes called L\'evy-Pareto distributions; their relevance in financial modelling has been known since the works of Mandelbrot and Fama in the 1960s \cite{Fama65,Mandelbrot63}. They are known to satisfy the space-fractional equation:
\begin{equation}\label{space_fractional}
\frac{\partial g_{\alpha}(x,t)}{\partial \tau} \, + \, \mu \, D^{\alpha} g_\alpha(x,t) = 0 
\end{equation}
where $D^\alpha:= \, ^{\alpha-2} D ^\alpha$ is a particular case of the Riesz-Feller operator \eqref{RF_def} for $\theta = \alpha-2$. The condition $\theta = \alpha -2$ turns out to be the fractional analogue to the probabilistic condition $\beta = -1$. Note that equation \eqref{space_fractional} degenerates into the the reduced BS equation \eqref{heat} when $\alpha=2$; the corresponding call option price is then
\begin{equation}\label{Levy_GreenForm} 
V_{\alpha}(S,K,r,\mu,\tau) \, = \, e^{-r\tau} \, \int\limits_{-\infty}^{+\infty} [Se^{(r+\mu)\tau+y}-K]^{+} \, g_\alpha(y,\tau) \, \id y
\end{equation}
where the risk-neutral parameter $\mu$ follows from \eqref{Levy_moment}:
\begin{equation}\label{Levy_mu}
\mu \, = \, \frac{\left( \frac{\sigma}{\sqrt{2}} \right)^\alpha}{\cos\frac{\pi\alpha}{2}}
\end{equation}
and reduces to $\mu=-\frac{\sigma^2}{2}$ in the Gaussian case ($\alpha=2$). An analytic resolution of the FMLS model has been provided in \cite{ACK17-Levy}, under the form of a quickly convergent series representation for the call price \eqref{Levy_GreenForm}. The proof is based on the Mellin-Barnes representation for the solutions of the space fractional equation \eqref{space_fractional} (see \cite{Mainardi01}): if $x>0$ then
\begin{equation}\label{Mellin_Green_LS}
g_{\alpha}(x,\tau) \, = \, \frac{1}{\alpha x } \, \int\limits_{c_1-i\infty }^{c_1+ i \infty} \,
\frac{\Gamma(1-t_1)}{\Gamma(1-\frac{t_1}{\alpha})}
\, \left( \frac{x}{(-\mu\tau)^{\frac{1}{\alpha}}} \right)^{t_1} \, \frac{\id t_1}{2i\pi}  \hspace*{1cm}  0 < c_1 < 1
\end{equation}

\subsection{Space-time option pricing model}

%\subsection{Option pricing under double fractional diffusion}

Let us discuss option pricing models based on space-time (double)-fractional diffusion equation, which can be expressed as
\begin{equation}\label{double_fractional}
\left({}^\ast_0 \mathcal{D}^\gamma_t +  \mu [{}^\theta \mathcal{D}^\alpha_x] \right) g(x,t) = 0\,
\end{equation}
where  $\alpha \in (0,2]$, $\gamma \in (0,\alpha]$. Asymmetry parameter $\theta$ is defined in the so-called \emph{Feller-Takayasu diamond} $|\theta| \leq \min \left\{\alpha, 2-\alpha \right\}$.  ${}^\ast_0 \mathcal{D}^\gamma_t$ denotes the \emph{Caputo fractional derivative}, which is defined as
\begin{equation}
{}^\ast_{t_0} \mathcal{D}^\nu_t f(t) =
\frac{1}{\Gamma(\lceil \nu \rceil - \nu)} \int_{t_0}^t \frac{f^{\lceil \nu \rceil}(\tau)}{(t - \tau)^{ \nu +1-\lceil \nu \rceil}} \ud \tau \end{equation}
and ${}^\theta \mathcal{D}^\alpha_x$ denotes the \emph{Riesz-Feller fractional derivative}, which is usually defined via its Fourier image as
\begin{equation}\label{RF_def}
\mathcal{F}[{}^\theta \mathcal{D}^\nu_x
f(x)](k) = - {}^\theta \psi^\nu (k)F[f(x)](k) = - \mu |k|^\nu e^{i(\mathrm{sign} k) \theta \pi /2} \mathcal{F}[f(x)](k)
\end{equation}

\noindent 
Let us describe the various financial models that are included in \eqref{double_fractional}.
The FMLS model, although far more generic than the BS one, can still be regarded as too restrictive; this is because the maximal negative asymmetry hypothesis $\beta = -1$, or equivalently $\theta = \alpha - 2$, does not describe well all capital markets (in particular illiquid ones, where financial assets often exhibit an almost symmetric heavy-tail). Nevertheless, it is not a priori possible to relax the maximal negative asymmetry hypothesis, because when $\beta\neq-1$ the expectations \eqref{mu_def} are known to diverge \cite{Carr03}. The fact that the risk-neutral parameter is infinite in this case traduces the fact that the risk cannot completely be eliminated from this class of L\'evy processes. Risk-minimal (instead of risk-neutral) approach has been introduced (see \cite{Bouchaud_Sornette94}) in this case; an interesting possibility, to generalize the the FMLS model and to remain within the risk-neutral framework, is to allow the time derivative to be also fractional (in the Caputo sense) in eq. \eqref{space_fractional}:
\begin{equation}\label{space_time_fractional}
^\ast_0 \mathcal{D}^\gamma_t g_{\alpha,\gamma}(x,t) \, + \, \mu_\gamma \ [{}^{2-\alpha}\, D^{\alpha}] g_{\alpha,\gamma}(x,t) = 0 
\end{equation}
The corresponding call option price now reads
\begin{equation}\label{DF_GreenForm}
V_{\alpha,\gamma}(S,K,r,\mu_\gamma,\tau) \, = \, e^{-r\tau} \, \int\limits_{-\infty}^{+\infty} [Se^{(r+\mu_\gamma)\tau+y}-K]^{+} \, g_{\alpha,\gamma}(y,\tau) \, \id y
\end{equation}
The Green functions are also known under the form of a Mellin-Barnes line integral \cite{Mainardi01}:
\begin{equation}\label{DF_Green_LS}
g_{\alpha,\gamma}(x,\tau) \, = \, \frac{1}{\alpha x } \, \int\limits_{c_1-i\infty }^{c_1+ i \infty} \,
\frac{\Gamma(1-t_1)}{\Gamma(1-\frac{\gamma}{\alpha}t_1)}
\, \left( \frac{x}{(-\mu_\gamma\tau)^{\frac{1}{\alpha}}} \right)^{t_1} \, \frac{\id t_1}{2i\pi}  \hspace*{1cm}  0 < c_1 < 1
\end{equation}
for any $x>0$. The main difference with the L\'evy-stable price \eqref{Levy_GreenForm} is that the risk-neutral parameter $\mu_\gamma$ now depends on the time-fractionality $\gamma$, and is not known analytically like in the L\'evy stable case \eqref{Levy_mu}. In \cite{ACK17}, an efficient and simple series expansion is derived for the risk neutral parameter, as well as a fast converging series expansion for the call price \eqref{DF_GreenForm}. In the next section, we discuss these results in detail, and test them in the real market conditions.

\section{Series representation of the pricing formulas under the space-time fractional diffusion}

Now, let us provide an analytic pricing formulas for the call options driven by the fractional diffusion \eqref{space_time_fractional} (details of the proofs can be found in \cite{ACK17}). We assume that $1<\alpha \leq 2$ and $0< \gamma \leq \alpha$. 

\subsection{Risk-neutral parameter}

The expectations in definition \eqref{mu_def} over the probability measure $\mathbb{P}$ can be expressed in terms of its probability densities $g_{\alpha,\gamma}(y,\tau)$, that is:
\begin{equation}\label{mu_density}
\mu_\gamma \, =\, -\log\int\limits_{-\infty}^\infty e^y g_{\alpha,\gamma}(y,\tau=1) \, \ud y
\end{equation}
It is possible to bring the calculation back to the non time-fractional case, by writing (see details in \cite{Zatloukal14}):
\begin{equation}\label{smearing}
g_{\alpha,\gamma}(x,\tau) = \int\limits_0^\infty g_\gamma(\tau,l) g_\alpha(l,x)\, \ud l
\end{equation}
where $g_\gamma$ and $g_{\alpha}^\theta$ are solutions of single-fractional diffusion equations
\begin{eqnarray}\label{single_fractional}
\frac{\partial g_\gamma(t,l)}{\partial l} &=& {}^\ast_0 \mathcal{D}^\gamma_t \, g_\gamma(t,l)\, \\
\frac{\partial g_\alpha(l,x)}{\partial l} &=&
 \mathcal{D}^\alpha_x \, g_\alpha(l,x)\, 
\end{eqnarray}

\subsubsection{Mellin-Barnes representation of the risk-neutral parameter}
The solution to the Caputo equation \eqref{single_fractional} is known to be \cite{Gorenflo99}:
\begin{equation}
g_\gamma(\tau,l) \, = \, \frac{1}{\tau^\gamma} \, M_\gamma\left( \frac{l}{\tau^\gamma} \right)
\end{equation}
where $M_\nu(z)$ is a function of Wright type, admitting the following Mellin-Barnes representation \cite{Mainardi10}:
\begin{equation}\label{Wright}
M_\nu(z) \,  =  \, \int\limits_{c-i\infty}^{c+i\infty} \, \frac{\Gamma(s)}{\Gamma(\nu s + 1 - \nu) } \, z^{-s} \, \frac{\ud s}{2i\pi} \hspace*{1cm} c>0\, .
\end{equation}
Inserting \eqref{smearing} and \eqref{Wright} in \eqref{mu_density}, interverting the integrals and using \eqref{Levy_moment} we obtain a Mellin-Barnes representation for the risk-neutral parameter:
\begin{equation}\label{mu_MB}
\mu_\gamma \, = \, - \log \left[ \frac{1}{\alpha} \, \int\limits_{c-i\infty}^{c+i\infty} \, \frac{\Gamma(s)\Gamma(\frac{1-s}{\alpha})}{\Gamma(\gamma s + 1 - \gamma) } \, \mu_{1}^{\frac{s-1}{\alpha}} \, \frac{\ud s}{2i\pi} \right] \hspace*{1cm} 0 < c < 1\, .
\end{equation}
where $\mu_1=\left( \frac{\sigma}{\sqrt{2}}\right) ^\alpha \sec{\frac{\pi \alpha}{2}}$ is the risk-neutral parameter in the L\'evy-stable case ($\gamma=1$), cf. eq. \eqref{Levy_mu}.

\subsubsection{Series representation of the risk-neutral parameter}
It is possible to express the integral over a vertical line \eqref{mu_MB} as a sum of residues of its analytic continuation, on the condition that the integrand decreases sufficiently fast at infinity. For Gamma function, this condition is determined by the well-known Stirling approximation \cite{Abramowitz72}
\begin{equation}\label{Stirling_complex_Gamma}
\Gamma(x \, + \, i y) \, \underset{\substack{|x| \rightarrow \infty \\ |y| \rightarrow \infty}}{\sim} \, \sqrt{2\pi} \, |y|^{x-\frac{1}{2}}\,e^{-\pi\frac{|y|}{2}}  \hspace*{1cm}  x,y \, \in \, \mathbb{R}  
\end{equation}
It follows from \eqref{Stirling_complex_Gamma} that, for a Gamma function of linear arguments of the type $\Gamma(a s +b), \, a,b\in\mathbb{R}$, its behavior at infinity depends on the sign of $a$, namely:
\begin{equation}\label{Gamma_linear}
\left\{
\begin{aligned}
& a > 0 \hspace*{1cm} |\Gamma(as+b)|\overset{|s|\rightarrow \infty}{\longrightarrow} 0  \hspace*{1cm}   \arg s \, \in \, (\frac{3\pi}{2} , -\frac{\pi}{2} )  \\
& a < 0 \hspace*{1cm} |\Gamma(as+b)|\overset{|s|\rightarrow \infty}{\longrightarrow} 0  \hspace*{1cm}   \arg s \, \in \, (-\frac{\pi}{2} , \frac{\pi}{2} )
\end{aligned}
\right.
\end{equation}
\eqref{Gamma_linear} easily generalizes to a ratio of products of Gamma functions of linear arguments. Let us assume that a function $f$ admits a Mellin transform $f^*$ of the form:
\begin{equation}
f^* (s) \, = \, \frac{\Gamma(a_1 s + b_1) \, \dots \, \Gamma(a_n s + b_n)}{\Gamma(c_1 s + d_1) \, \dots \, \Gamma(c_m s + d_m)} \hspace*{1cm} a_j, b_j, c_k, d_k \in \mathbb{R}
\end{equation}
and that this Mellin transform converges on some non-empty strip $\{ Re(s) \, \in \, (c_1,c_2)\}$, so that the Mellin inversion formula holds:
\begin{equation}\label{Mellin_inversion}
f(x) \, = \,  \int \limits_{c-i\infty}^{c+i \infty} \, f^*(s) \, x^{-s} \, \frac{\ud s}{2i\pi} \hspace*{1cm} c \in (c_1,c_2)
\end{equation}
Introduce the characteristic quantity $\Delta$:
\begin{equation}\label{Delta_1D}
\Delta \, = \, \sum\limits_{j=0}^n \, a_j \, - \, \sum\limits_{k=0}^m \, c_k
\end{equation}
It follows from \eqref{Gamma_linear} that $\Delta$ governs the asymptotic behavior of $f^*(s)$:
\begin{equation}\label{Gamma_linear_n}
\left\{
\begin{aligned}
& \Delta > 0 \hspace*{1cm} |f^*(s)| \overset{|s|\rightarrow \infty}{\longrightarrow} 0  \hspace*{1cm}  \arg s \, \in \, (\frac{3\pi}{2} , -\frac{\pi}{2} ) \\
& \Delta < 0 \hspace*{1cm} |f^*(s)| \overset{|s|\rightarrow \infty}{\longrightarrow} 0  \hspace*{1cm}  \arg s \, \in \, (-\frac{\pi}{2} , \frac{\pi}{2} )
\end{aligned}
\right.
\end{equation}
Therefore, applying the residue theorem to the inversion formula \eqref{Mellin_inversion} yields:
\begin{equation}\label{Gamma_residue_n}
\left\{
\begin{aligned}
& \Delta > 0 \hspace*{1cm} f(x) \, = \, \sum\limits_{Re(s) < c}  \res \left[ f^*(s)x^{-s} \right] \\
& \Delta < 0 \hspace*{1cm} f(x) \, = \, -\sum\limits_{Re(s) > c}  \res \left[ f^*(s)x^{-s} \right]
\end{aligned}
\right.
\end{equation}
where the choice $\{ Re(s) > c \}$ or $\{ Re(s) < c \}$ is determined by the fact that $|x|^{-s}$ goes to $0$ at infinity in the chosen half-plane. In the case of the  Mellin-Barnes representation \eqref{mu_MB}, the characteristic quantity is 
\begin{equation}
\Delta \, = \,  \, 1 - \frac{1}{\alpha} \, - \, \gamma
\end{equation}
and is negative as soon as:
\begin{equation}\label{cond_gamma}
\gamma \, > \, 1 - \frac{1}{\alpha}
\end{equation}
It follows from rule \eqref{Gamma_residue_n} that one can express \eqref{mu_MB} as the sum of the residues in the right half-plane (we choose it because $\mu_1^{\frac{s-1}{\alpha}}$ goes to 0 at infinity in this half plane as soon as $|\mu_1|<1$, which is the case in all financial applications). These poles are induced by the singularities of the $\Gamma(\frac{1-s}{\alpha})$ term, which arise at every negative integer value $-n$ of its argument, that is at every point of the type $s=1+\alpha n, \, n\in\mathbb{N}$; it is well-known the residue of the Gamma function at a negative integer $-n$ is $\frac{(-1)^n}{n!}$ \cite{Abramowitz72}, and therefore we obtain:
\begin{equation}\label{mu_series}
\mu_\gamma \, = \, - \log \, \sum\limits_{n=0}^{\infty} \, \frac{(-1)^n \Gamma(1+\alpha n)}{n!\Gamma(1+\gamma\alpha n)} \mu_1^n\, 
\end{equation}
as soon as the condition \eqref{cond_gamma} is fulfilled. An interesting approximation of \eqref{mu_series} can be easily derived from the Taylor approximation $\log(1+u) \simeq u$:
\begin{eqnarray}\label{mu_approx}
\mu_\gamma & = & -\log \left[  1 \, - \, \frac{ \Gamma(1+\alpha)}{\Gamma(1+\gamma\alpha )} \mu_1 \, + \, \dots \right] \nonumber \\
 & \simeq &  \frac{\Gamma(1+\alpha)}{\Gamma(1+\gamma\alpha )} \mu_1
\end{eqnarray}
which, as expected, coincides with the L\'evy-stable risk-neutral parameter $\mu_1$ when $\gamma=1$. As a particular case, we obtain a nice approximation for the risk-neutral parameter in the fractional Black-Scholes model ($\alpha = 2$):
\begin{equation}\label{mu_fBS_approx}
\mu_\gamma \, \simeq \, -\frac{\sigma^2}{\Gamma(1+2\gamma)}
\end{equation}
which resumes to the well-known gaussian parameter $-\frac{\sigma^2}{2}$ when $\gamma=1$.

\begin{figure}[t]
\centering
\includegraphics[scale=0.4]{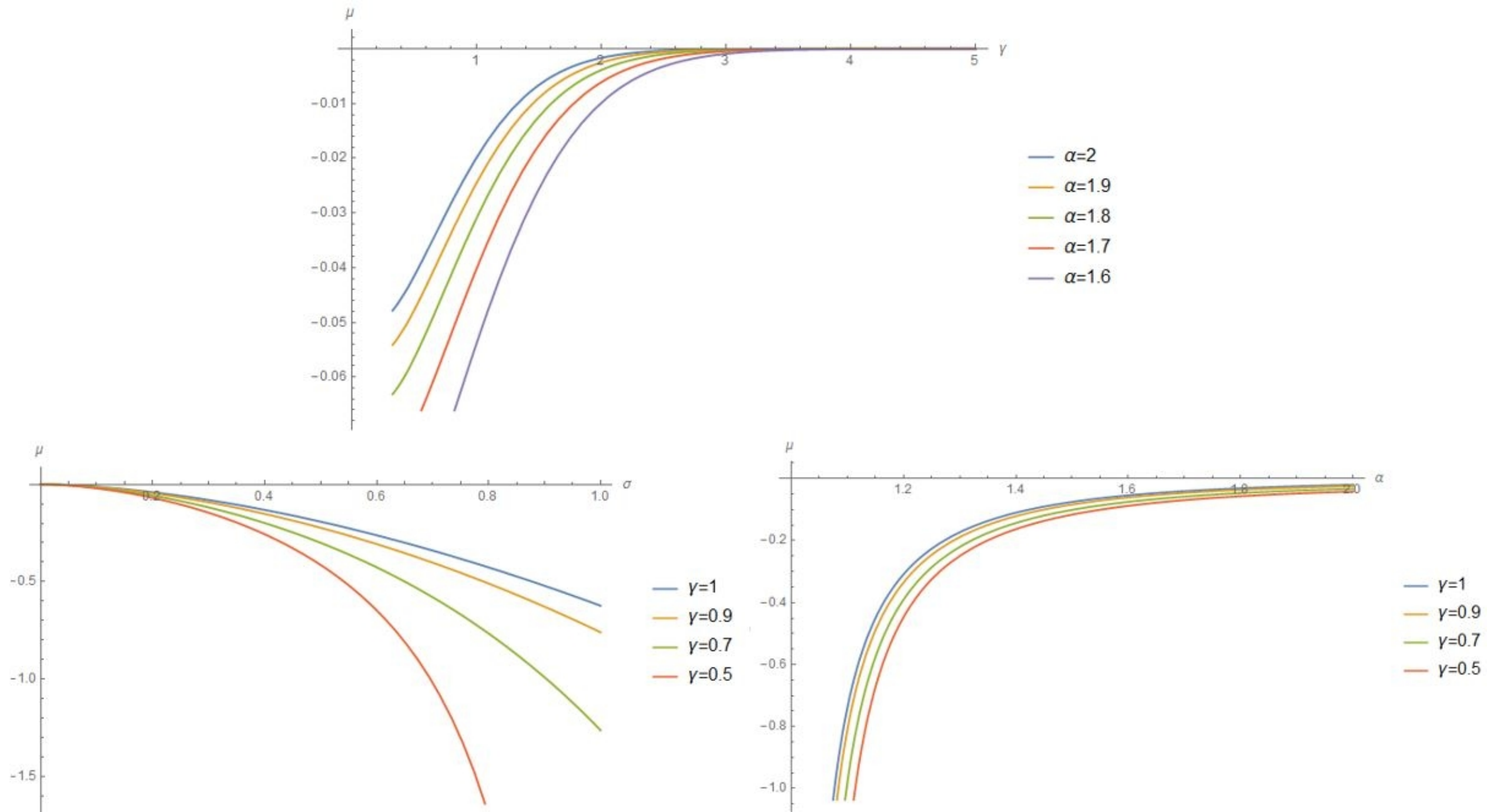}
\caption{In the first graph, we plot the evolution of $\mu_\gamma$ in function of $\gamma$ for different stability parameters $\alpha\in[1.6,1]$ and market volatility $\sigma = 20\%$. We only consider $\gamma>0.38$ so that the condition $\gamma>1-\frac{1}{\alpha}$ is fulfilled for any of the chosen stabilities. In graph 2 and 3, we plot the evolution of $\mu_\gamma$ in function of the market volatility and the stability parameter resp., for different values of the fractionality $\gamma$.}
\label{fig1}
\end{figure}

\noindent
In Fig. \ref{fig1}, we plot the evolution of $\mu$ in function of the parameters $\gamma$, $\sigma$ and $\alpha$; thanks to the exponential convergence of the series \eqref{mu_series}, it suffices to consider only the very first few terms of the series to obtain an excellent level of precision.
\noindent
% We may note that, when taking $\gamma = 1$ in formula \eqref{mu_series}, we are left with
% \begin{equation}
% \mu \, =  \, - \log \, \sum\limits_{n=0}^{\infty} \frac{(-1)^n\mu_1^n}{n!} \, = \, - \log \, e^{-\mu_1} \, = \, \mu_1
% \end{equation}
% as expected.

\subsection{Option price}
In all the following we will use the notation $[\log]:=\log\frac{S}{K}+r\tau$, so that the payoff in \eqref{DF_GreenForm} can be written:
\begin{equation}\label{payoff_log}
[Se^{(r+\mu_\gamma)\tau+y}-K]^+ \, = \, K[e^{[\log]+\mu_\gamma \tau+y}-1]^+
\end{equation}

\subsubsection{Mellin-Barnes representation of the option price}
The call price \eqref{DF_GreenForm} can be expressed as a double Mellin-Barnes integral. First, one introduces in \eqref{DF_GreenForm} the Mellin-Barnes representation \eqref{DF_Green_LS} for the Green function; second, one writes a Mellin-Barnes representation for the exponential term in \eqref{payoff_log}:
\begin{equation}
e^{[\log]+\mu_\gamma\tau + y} \, = \, \int\limits_{c_2-i\infty}^{c_2+i\infty} \, (-1)^{-t_2} \Gamma(t_2) ([\log] + \mu_\gamma\tau + y )^{-t_2} \, \frac{dt_2}{2i\pi} \hspace*{1cm} c_2 > 0\, 
\end{equation}
The integral over the Green variable $y$ becomes a particular case of a B\^eta integral, which is straightforward to calculate, and one obtains the representation for the option price:
\begin{multline}\label{Call_C2}
V_{\alpha,\gamma} (S,K,r,\mu_\gamma,\tau) \, = \, \frac{K e^{-r \tau}}{\alpha}
\int\limits_{c_1-i\infty}^{c_1+i\infty} \int\limits_{c_2-i\infty}^{c_2+i\infty}  \, (-1)^{-t_2}
\frac{\Gamma(t_2)\Gamma(1-t_2)\Gamma(-1-t_1+t_2)}{\Gamma(1-\frac{\gamma}{\alpha}t_1)} \, \\
\times
(-[\log]-\mu_\gamma\tau)^{1+t_1-t_2}(-\mu_\gamma\tau^{\gamma})^{-\frac{t_1}{\alpha}}\frac{\ud t_1}{2i\pi}\wedge\frac{\ud t_2}{2i\pi}\, 
\end{multline}
The vector $\underline{c}:= [ c_1 , c_2 ]$ is an element of the $\mathbb{C}^2$-polyhedra $P:=\{ (t_1,t_2)\in\mathbb{C}^2, \, 0<Re(t_2)<1 , \, Re(t_2-t_1)>0  \}$, which generalizes the notion of convergence strip for one-dimensional Mellin transform.

\subsubsection{Series representation of the option price}
\begin{figure}[t]
\centering
\includegraphics[scale=0.4]{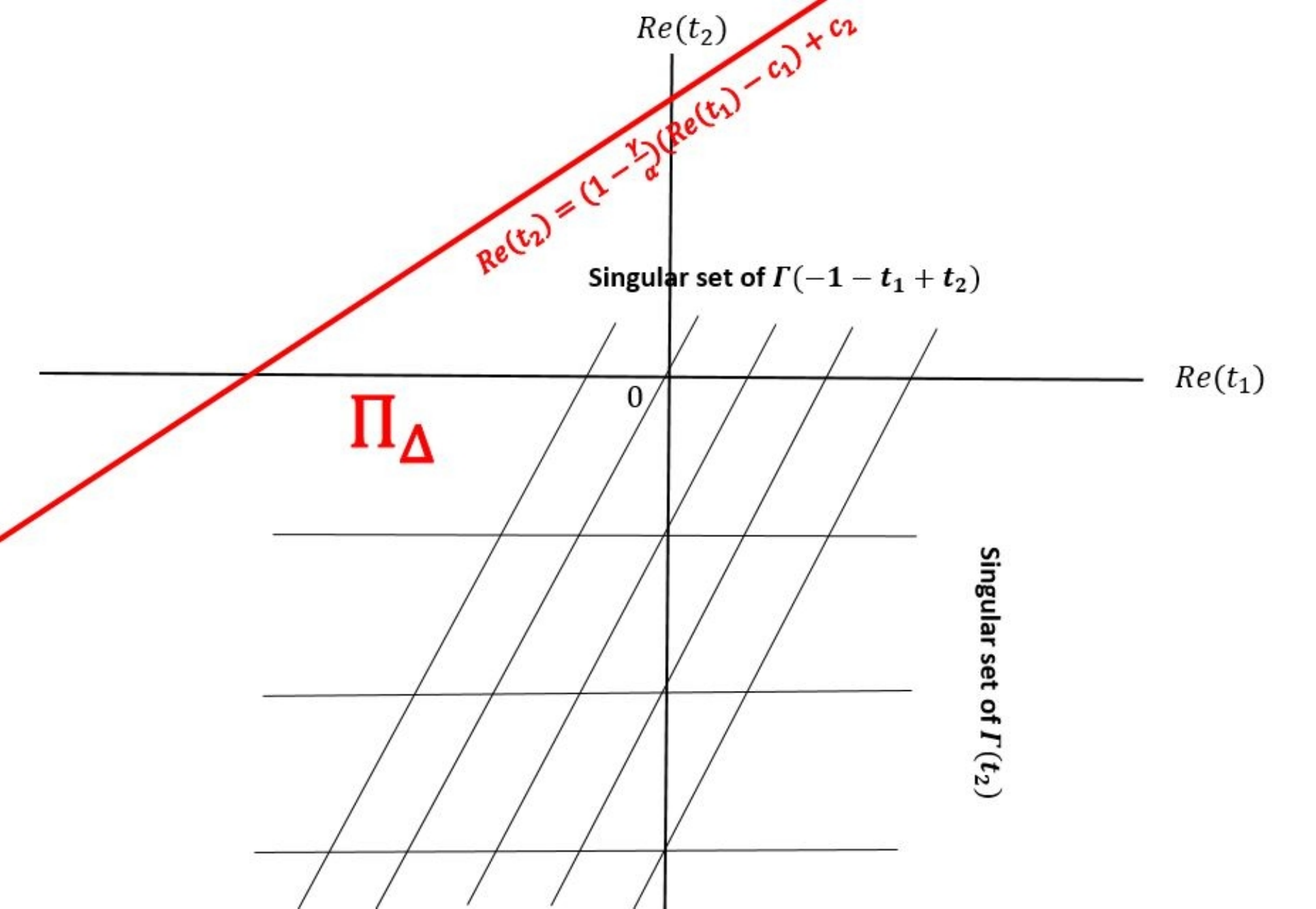}
\caption{Residues contributing to the evaluation of the double Mellin-Barnes integral \eqref{Call_C2}.}
\label{fig-divisors}
\end{figure}

The double Mellin-Barnes integral \eqref{Call_C2} can also be expressed as a sum of residues. In one dimension, it is usual to sum the residues right or left to the convergence strip of the Mellin transform, like we have done for the risk-neutral parameter, where the integral \eqref{mu_MB} has been computed by right-summing the residues to obtain the series \eqref{mu_series}. In two dimensions, this procedure generalizes to a summation to a subregion of $\mathbb{C}^2$, determined by a characteristic vector associated to the integrand. The incoming residues are computed by the two-dimensional analogue to the Cauchy formula:
\begin{equation}\label{Cauchy2D}
\res \left[ f(t_1,t_2) \, \frac{\ud t_1}{2i\pi t_1} \wedge \frac{\ud t_2}{2i\pi t_2} \right] \, = \, f(0,0)
\end{equation}
This procedure has been introduced in \cite{Tsikh94,Tsikh97}. Namely, the characteristic quantity \eqref{Delta_1D} generalizes to a characteristic vector, which, in the case of the double Mellin-Barnes integral \eqref{Call_C2} reads
\begin{equation}\label{Delta_2D}
\Delta \, = \, 
\begin{bmatrix}
-1 + \frac{\gamma}{\alpha} \\
1
\end{bmatrix}
\end{equation}
The rule \eqref{Gamma_residue_n} generalizes to
\begin{equation}
f(x_1,x_2) \, = \, \sum\limits_{[t_1,t_2]\in\Pi_\Delta} \, \res \left[ f^*(t_1,t_2) \, x_1^{-t_1} \, x_2^{-t_2} \, \right]
\end{equation}
where $\Pi_\Delta$ is the subset of $\mathbb{C}^2$ defined by
\begin{equation}
\Pi_\Delta \, = \, \left\{  \underline{t}:=[t_1,t_2]\in\mathbb{C}^2 \, ,\, Re( \Delta . \underline{s} ) \, < \, Re ( \Delta . \underline{c} )  \right\}
\end{equation}
in the sense of the euclidean scalar product. In the plane $Re  \left( \mathbb{C}^2 \right)$, $\Pi_\Delta$ is therefore the region located under the line
\begin{equation}
Re(t_2) \, = \, (1 - \frac{\gamma}{\alpha} ) (Re(t_1) - c_1) \, + \, c_2
\end{equation}
whose slope is positive because by hypothesis $\gamma \leq \alpha$. In this region, poles come from functions $\Gamma(-1-t_1+t_2)$ and $\Gamma(t_2)$ which are singular at every negative integer value of their argument (see Fig.~\ref{fig-divisors}).

\noindent 
From the singular behavior of the Gamma function around a singularity \cite{Abramowitz72} and the Cauchy formula \eqref{Cauchy2D}, we obtain the series for the call price under double-fractional model:
\begin{equation}\label{Formula}
V_{\alpha,\gamma} (S,K,r,\mu_\gamma,\tau) \, = \, \frac{Ke^{-r\tau}}{\alpha} \, \sum\limits_{\substack{n = 0 \\ m = 1}}^{\infty} \, \frac{(-1)^n}{n!\Gamma(1-\gamma\frac{n-m}{\alpha})} (-[\log]-\mu_\gamma\tau)^{n}(-\mu_\gamma\tau^{\gamma})^{\frac{m-n}{\alpha}}\, 
\end{equation}

\noindent
Full details of this calculation can be found in \cite{ACK17}.

\section{Applications}

Let us discuss several applications of the series formula for the space-time fractional option prices. We show that it can be used for estimating the market parameters of the option prices. The calculation of the option price is very quick compared to the other methods (Mellin-Barnes representation, numerical estimation, \dots). We also briefly discuss the applications to implied volatility. 

\subsection{Call price}

\begin{figure}[h]
\centering
\includegraphics[scale=0.3]{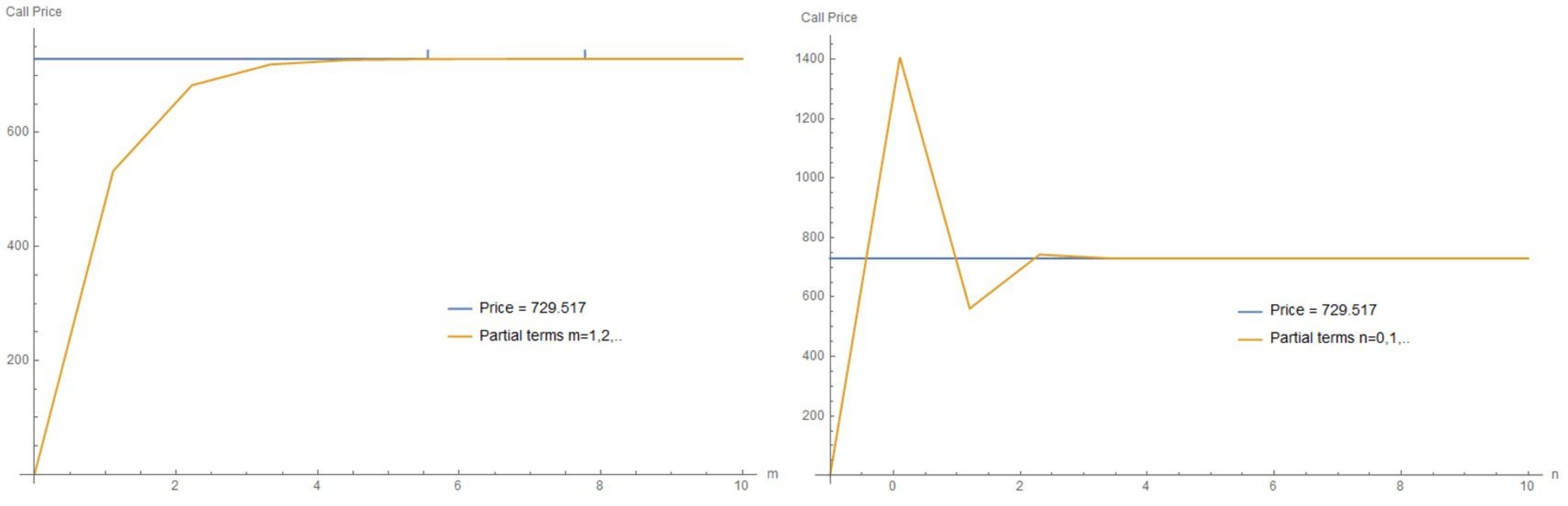}
\caption{Convergence of the $m$ and $n$ partial sums of the double-fractional call price series.}
\label{fig_partial_sums}
\end{figure}

\noindent
When fixing an upper bound for the $n$ (resp. $m$) summation in the double series \eqref{Formula}, we are left with a simple series whose $m$-(resp. $n$) partial sums converges very quickly to the option price (see fig. ~\ref{fig_partial_sums}, where the parameters are $S=3800,K=4000,r=1\%,\sigma=20\%, \alpha=1.7, \gamma = 0.9$). We may observe that the convergence of the $m$-sums are monotone, while the $n$-sums oscillate around the final price.

\begin{figure}[h]
\centering
\includegraphics[scale=0.5]{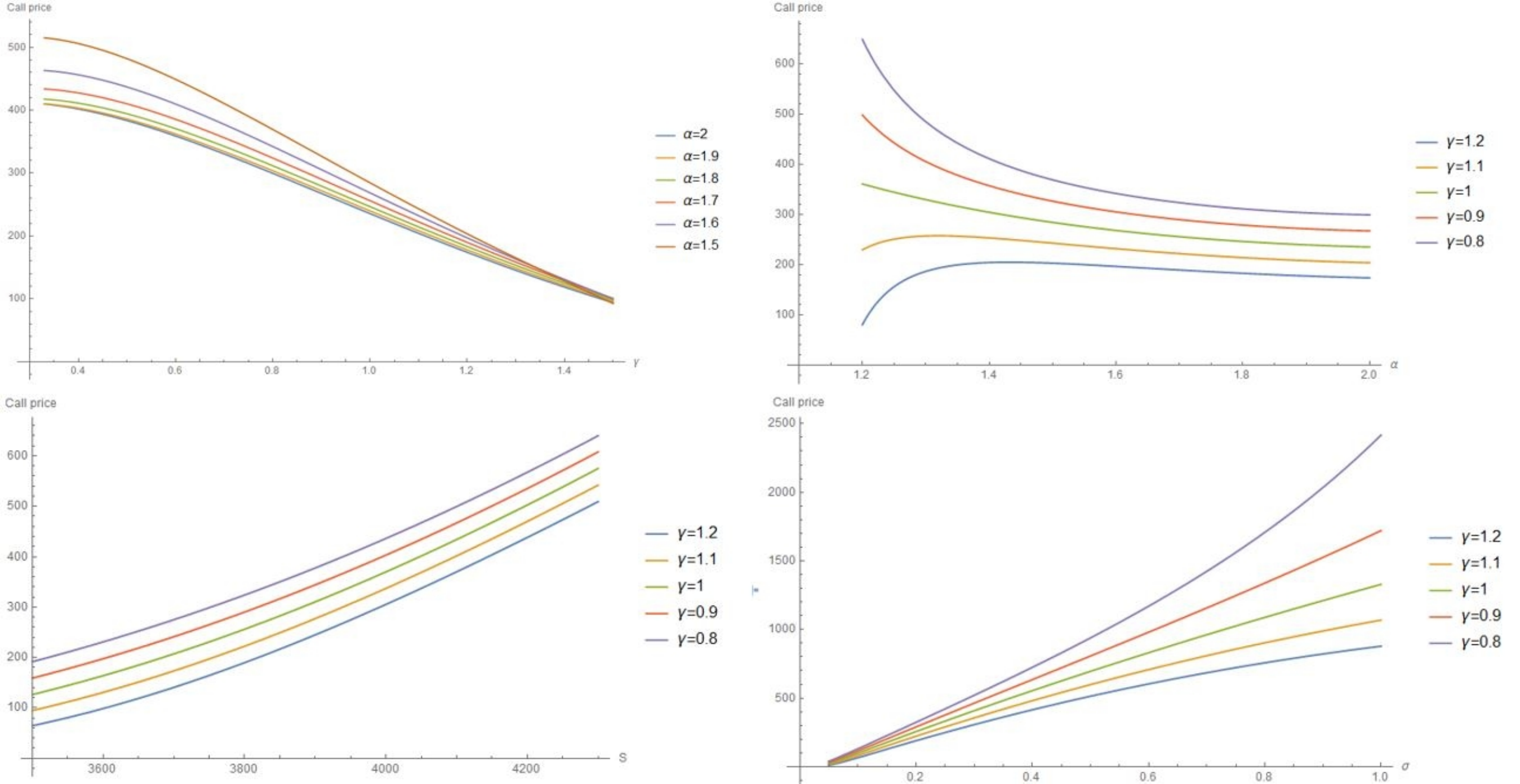}
\caption{Evolution of the double-fractional call price in function of various parameters (time-fractionality parameter $\gamma$, stability parameter $\alpha$, asset (spot) price $S$ and market volatility $\sigma$ .}
\label{fig2}
\end{figure}

\noindent
In the graphs in fig. \ref{fig2}, we study the evolution of the option price \eqref{Formula} in function of different parameters. In the first graph we fix $S=3800$, $K=4000$, r=1\% $\sigma= 20\%$ and we plot the evolution of the price in function of $\gamma$, for different stability parameters $\alpha\in[1.5,2]$; we choose to consider only $\gamma>0.33$ so that the condition $\gamma>1-\frac{1}{\alpha}$ is satisfied for all stabilities, and observe that the prices are a decreasing function of the time fractionality. In graph 2 we let $\alpha$ vary between $1$ and $2$ and note that when $\gamma\leq1$ then the prices are always a decreasing function of the stability, while for $\gamma>1$ they possess a maximum. In graph 3 (resp. 4) we plot the evolution of the option price in function of the spot price $S$ (resp. market volatility $\sigma$) for various time fractionality $\gamma$, and with fixed stability $\alpha=1.7$; note that the prices are, as expected, always a monotonous (growing) function of the spot and of the volatility, which is coherent with the non-arbitrage hypothesis of financial markets. Finally, we show the estimated parameters of the three option pricing models for the real options of S\&P 500 options. The results are presented in Tab. \ref{tab:est}, which has been taken from \cite{KK16}.

\begin{table}[t]
\centering
\begin{tabular}{|c|ccc|}
  \hline
  \multicolumn{4}{|c|} {All options}\\
  \hline
 parameter& Black-Scholes & L\'{e}vy stable & Double-fractional  \\
\hline
  $\alpha$ & - & 1.493(0.028)& 1.503(0.037)\\
  $\gamma$  & - & - & 1.017(0.019) \\
  $\sigma$ & 0.1696(0.027)& 0.140(0.021)& 0.143(0.030) \\
   AE & 8240(638)& 6994(545)& 6931(553)\\
  \hline
  \hline
  \multicolumn{4}{|c|} {Call options}\\
  \hline
 parameter& Black-Scholes & L\'{e}vy stable & Double-fractional  \\
\hline
  $\alpha$ & - & 1.563(0.041)& 1.585(0.038)\\
  $\gamma$  & - & - & 1.034(0.024) \\
  $\sigma$ & 0.140(0.021) & 0.118(0.026) & 0.137(0.020)  \\
   AE & 3882(807) & 3610(812) & 3550(828) \\
  \hline
  \hline
    \multicolumn{4}{|c|} {Put options}\\
  \hline
 parameter& Black-Scholes & L\'{e}vy stable & Double-fractional  \\
\hline
  $\alpha$ & - & 1.493(0.031)& 1.508(0.036)\\
  $\gamma$ & - & - & 1.047(0.017)\\
  $\sigma$ & 0.193(0.039) & 0.163(0.034) & 0.163(0.037)   \\
  AE & 3741(711) & 3114(591) & 2968(594)\\
  \hline
\end{tabular}
\caption{Estimated values of option pricing parameters based on Black-Scholes model, FMLS model and Space-time fractional model. The estimation was done for all options and separately for call options and put options, respectively. We see that for this case is $\gamma$ very close to one, which does not have to be true for illiquid markets or during the abnormal periods. $AE$ denotes the aggregated error, which is defined as the sum of absolute differences between estimated price and market price. Table was taken from Ref. \cite{KK16} and it is possible to find more details about the estimation there.}
\label{tab:est}
\end{table}

\subsection{Implied volatility}

The process of implying the market volatility consists in finding for which volatility $\sigma_I$ a model-driven option price coincides with the observable price $C$, that is when
\begin{equation}\label{implied}
V(S,K,r,\sigma_I,\tau) \, = \, C
\end{equation}
A typical procedure is to imply a Black-Scholes volatility (by using the Black-Scholes formula for the price and solving \eqref{implied} by means of numerical methods, such as a Newton-Raphson algorithm, see for instance \cite{Wilmott06}) and use it as an input parameter in a more sophisticated model. Let us show how the analytic series \eqref{Formula} allows to imply a market volatility, and compare with the Gaussian one.

\subsubsection{At-the-money volatility}

\begin{figure}[h]
\centering
\includegraphics[scale=0.5]{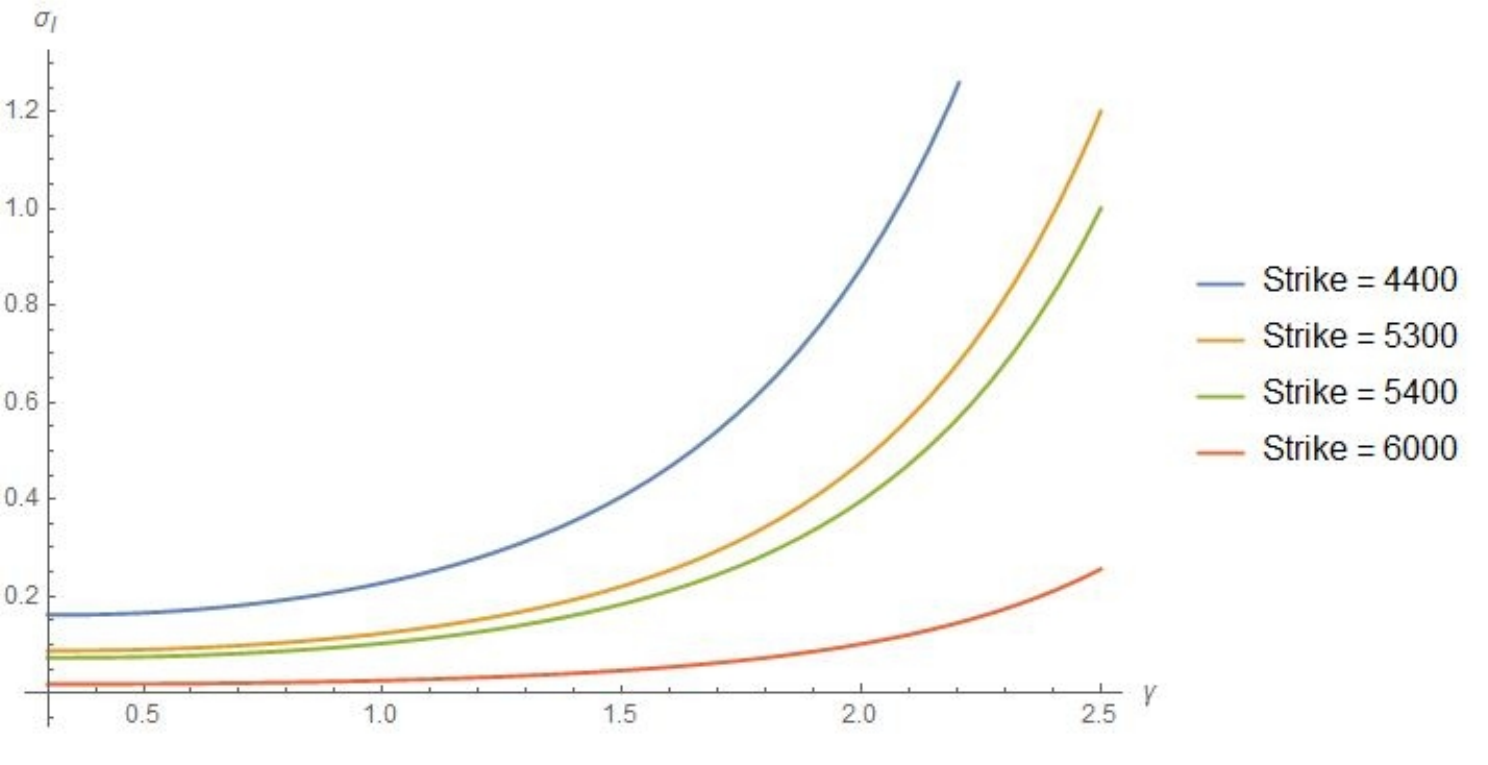}
\caption{The at-the-money implied volatility for the double-fractional Black-Scholes model, as a function of time fractionarity $\gamma$.}
\label{fig:fBS-ATMfwvol}
\end{figure}

\noindent When the asset is "at-the-money forward", that is when
\begin{equation}\label{ATMfwd}
S \, = \, K e^{-r\tau}
\end{equation}
then there exists an approximation for the Black-Scholes formula \cite{BS94}
\begin{equation}
V(S,K,r,\sigma,\tau) \, \simeq \, \frac{S}{\sqrt{2\pi}} \sigma \sqrt{\tau}
\end{equation}
and therefore the solution to the implied volatility equation \eqref{implied} reads
\begin{equation}\label{BS_implied}
\sigma_I \, \simeq \, \frac{C}{S} \, \sqrt{\frac{2\pi}{\tau}}
\end{equation}
Such an approximation can also be derived in the double-fractional Black-Scholes model ($\alpha=2$): note that, with our notations, the ATM-forward hypothesis \eqref{ATMfwd} reads $[\log]=0$ and therefore in this case the pricing formula \eqref{Formula} becomes a power series (i.e., with only positive powers of $\mu_\gamma$ and $\tau$):
\begin{equation}\label{Formula_approx}
V_{2,\gamma}(S,K,r,\mu,\tau) \, = \, \frac{S}{2} \, \left[ \, \frac{1}{\Gamma(1+\frac{\gamma}{2})} \sqrt{-\mu_\gamma \tau^\gamma}  \, + \, O(-\mu\tau^\gamma) \, \right]
\end{equation}
Using approximation \eqref{mu_fBS_approx} for the risk-neutral parameter
\begin{equation}
\mu_\gamma \, = \, - \frac{\sigma^2}{\Gamma(1+2\gamma)}
\end{equation}
in the first order term of the power series \eqref{Formula_approx}, we obtain the implied fractional Black-Scholes volatility (in the ATM forward case):
\begin{equation}\label{DFBS_implied}
\sigma_I \, \simeq \, 2 \, \frac{C}{S}  \, \Gamma(1+\frac{\gamma}{2}) \, \sqrt{\frac{\Gamma(1+2\gamma)}{\tau^\gamma}}
\end{equation}
Let us remark that the formula \eqref{DFBS_implied} resumes to the Black-Scholes implied volatility formula \eqref{BS_implied} when $\gamma=1$ (recall that $\Gamma(\frac{3}{2}) = \frac{\sqrt{\pi}}{2}$). In graph \ref{fig:fBS-ATMfwvol} we plot the evolution of formula \eqref{DFBS_implied} in function of $\gamma$ for a time to maturity $\tau = 1.027$ and various exercise and call prices (see market datas in table~\ref{fig:volsmile_data}).

\subsubsection{Volatility smile}

% -- I'M NOT SUR OF THIS SUBSUBSECTION. FOR THE MOMENT I DONT SEE HOW TO APPLY A NEWTON RAPHSON ALGORITHM TO THE POWER SERIES OF THE CALL PRICE (IN ORDER TO SOLVE THE EQUATION FOR IMPLICIT VOL). IN TABLE 1 I SIMPLY USED A MATEHATMICA "SOLVE" TO THE SERIES TRUNCATED TO N,M = 4 BUT I DOUBT THE RESULTS ARE TRUe. HOW TO MAKE IT CLEAN?

In table \ref{fig:volsmile_data}, we provide observable market bid (offered) prices for S \& P 500  call options with several exercise (strike) prices traded end 2008, quotation date = 03 nov. 2008, expiry = 17 jan. 2009 (source: eurexchange.com). We compute the implied Black-Scholes volatility as well as the implied fractional Black-Scholes volatility for various time fractionalities. They are obtained via a truncation of the series \eqref{Formula} to $n,m=4$ and the approximation \eqref{mu_fBS_approx} for the parameter $\mu_\gamma$.

\begin{table}[h!]
\centering
\begin{scriptsize}
\begin{tabular}{|cc||cccc|}
  \hline
 Strike  &  Call price & BS vol & f-BS vol ($\gamma=0.8$) & f-BS vol ($\gamma=0.9$) & f-BS vol ($\gamma=1.1$)   \\
  \hline
  \hline
  900 & 118.9 & 0.4708 & 0.3163 & 0.3827 &  0.5900         \\
  940 & 92.7  & 0.4462 & 0.3066 & 0.3670  & 0.5330         \\
  980 & 69.5  & 0.4232 & 0.2929 &  0.3493 & 0.5210        \\
  1020 & 49.2 & 0.3976 & 0.2754 & 0.3284  & 0.4891         \\
  1060 & 32.3 & 0.3711 & 0.2557 & 0.3058 &  0.4574  \\
  1100 & 19.5 & 0.3475 & 0.2380 & 0.2857 & 0.4186 \\
  1150 & 8.9 & 0.3279 & 0.2269 & 0.2727 & 0.3938 \\
  1180 & 5.1 & 0.3301 & 0.2324 & 0.2789 & 0.3764 \\
  1220 & 2 & 0.3514 & 0.2514 & 0.3015 & 0.3692 \\
  1280 & 0.25 & 0.4110 & 0.2949 & 0.3544 & 0.4166 \\
  \hline
\end{tabular}
\end{scriptsize}
\caption{Implied volatility for S\& P 500 index call options}
\label{fig:volsmile_data}
\end{table}

\begin{figure}[h!]
\centering
\includegraphics[scale=0.5]{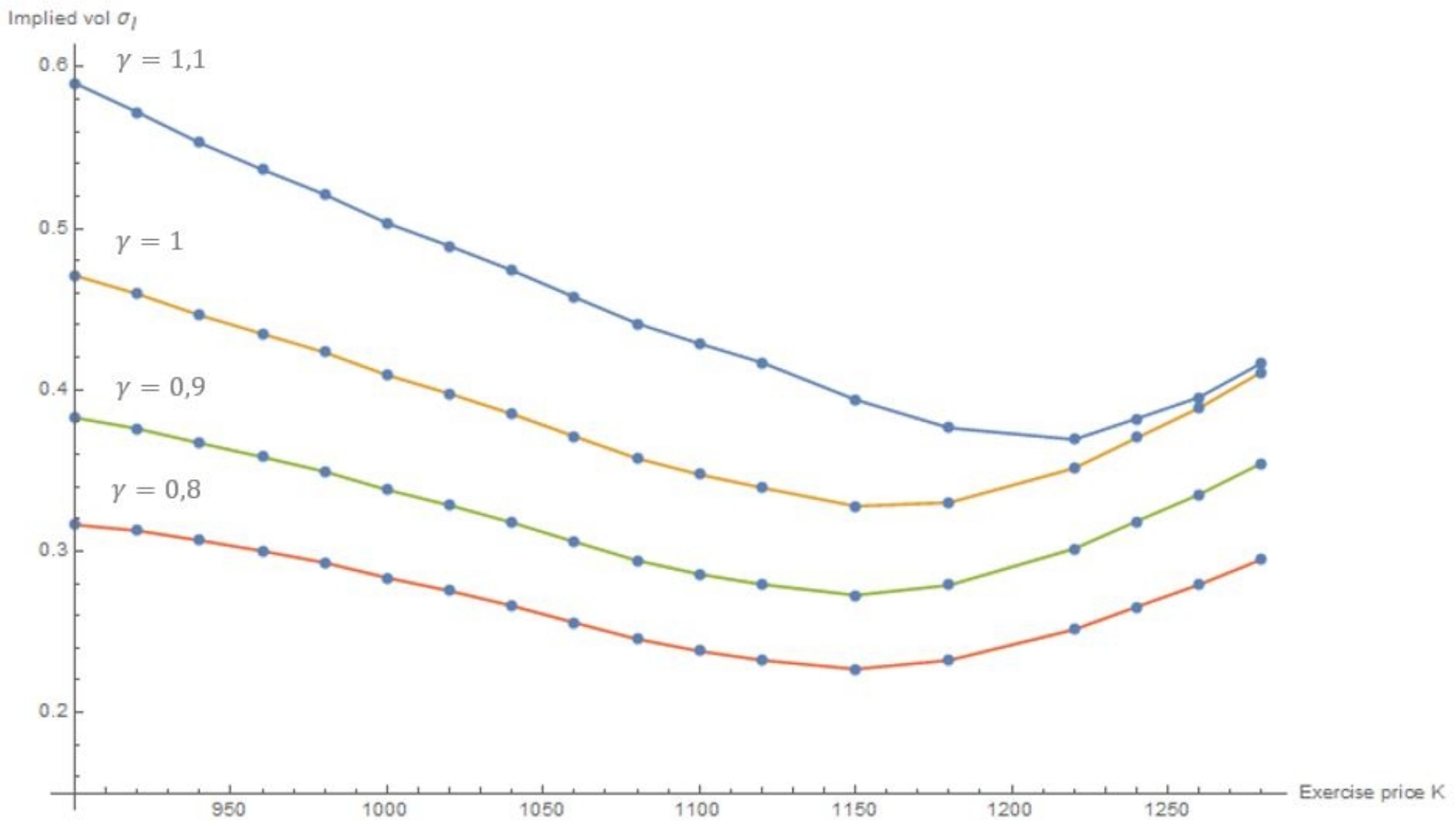}
\caption{Implied volatility for the fractional Black-Scholes model, (market price S = 966.3)}
\label{fig:ImpliedVol1}
\end{figure}

\noindent In fig \ref{fig:ImpliedVol1} we plot the implied volatilities obtained in table \ref{fig:volsmile_data} for and for $0.8 \leq \gamma \leq 1.1$. We observe that the usual volatility smile (that is, the existence of a minimum around the spot price) is preserved, although less smooth when $\gamma>1$. Interestingly, when $\gamma \leq 1$, the minimal implied volatility is attained for the same strike price (independently of $\gamma$).

% \begin{figure}[h!]
% \centering
% \includegraphics[scale=0.4]{ImpliedVol1.pdf}
% \caption{Implied volatility for the fractional Black-Scholes model, for $\gamma\leq 1$ (market price S = 5395.98}
% \label{fig:ImpliedVol1}
% \end{figure}

% \begin{figure}[h]
% \centering
% \includegraphics[scale=0.5]{ATMvol.pdf}
% \caption{The at-the-money implied volatility for the double-fractional model, as a function of fractionality and of stability.}
% \label{fig:ATMfwvol}
% \end{figure}

%%%%%%%%

% \begin{figure}[h]
% \centering
% \includegraphics[scale=0.4]{convergence.pdf}
% \caption{}
% \label{fig1}
% \end{figure}

\section{Conclusions}
In this paper, we have discussed the application of space-time fractional diffusion in option pricing and its relation to Black-Scholes model and Finite moments L\'{e}vy stable model. Models based on fractional diffusion enable to model the risk redistribution in order to incorporate large drops, memory effects and abnormal periods. We have briefly introduced all aforementioned models and described their main properties. Additionally, we have presented the series representation for all models, which is based on Mellin-Barnes integral representation of the option price and residue summation in $\mathbb{C}^2$. This mathematical techniques can overcome the technical difficulties of the fractional models, which is caused by the fact that the resulting prices are normally expressed in terms of integral transforms (Fourier, Laplace, Mellin) and the practical calculation is time consuming and understandable only to people trained in fractional calculus. The resulting series representation can be easily grasped by any financial practitioner. We have also applied the formulas to real financial data in order to demonstrate fast convergence and stability of the method. We have particularly shown numerical estimations of model parameters from the real data, applications to implied volatility and presence of volatility smile. 

Fractional models provide a fruitful field for further investigations of  financial systems, including portfolio management, derivative pricing, commodity pricing and many other possible applications. Naturally, in these applications it is necessary to carefully define the proper fractional derivatives and boundary conditions. In some cases, as e.g. in the case of fractional geometric Brownian motion, it is also necessary to overcome the mathematical issues, as non-existence of moments, etc.  Some of these topics will be addressed in the future research.  

\section*{Acknowledgements}
J. K. acknowledges support from the Austrian Science Fund, Grant No. I 3073-N32., and from the Czech Science Foundation, Grant No. 17–33812L.

\end{document}